\documentclass[12pt,preprint]{aastex}
\begin{document}
\def\dd{\partial}
\def\lta{\mathrel{\rlap{\lower 3pt\hbox{$\mathchar"218$}}
    \raise 2.0pt\hbox{$<$}}}

\shorttitle{Nonradiative Accretion Flows}
\shortauthors{Hawley \& Balbus}

\title{The Dynamical Structure of Nonradiative Black Hole Accretion Flows}

\author{John F. Hawley\altaffilmark{1} 
and Steven A. Balbus\altaffilmark{1}}

\altaffiltext{1}{Dept. of Astronomy, University of Virginia, PO Box
3818, Charlottesville,\\ VA 22903, USA. jh8h@virginia.edu;
sb@virginia.edu}

\begin{abstract}

We analyze three-dimensional magnetohydrodynamic (MHD) simulations of a
nonradiative accretion flow around a black hole using a
pseudo-Newtonian potential.  The flow originates from a torus
initially centered at 100 gravitational (Schwarzschild) radii.
Accretion is driven by turbulent stresses generated self-consistently
by the magnetorotational instability.  The resulting flow has three
well-defined dynamical components:  a hot, thick,
rotationally-dominated Keplerian disk; a surrounding magnetized corona
with vigorous circulation and outflow; and a magnetically-confined jet
along the centrifugal funnel wall.  Inside of 10 gravitational radii,
the disk becomes very hot, more toroidal, and highly intermittent.
These results contrast sharply with quasi-spherical, self-similar viscous
models.  There are no significant dynamical differences between
simulations that include resistive heating and those that do not.  We
conclude by deducing some simple radiative properties of our solutions,
and apply the results to the accretion-powered Galactic center source
Sgr A*.

\end{abstract}


\section {Introduction}

Many black hole accretion sources are remarkably underluminous, and
pose a stern challenge for accretion theory.  Generally, these sources
emit a significant fraction of their radiated energy in the form of
X-rays (Elvis et al.~1994), and standard, optically thick Keplerian
disk theory cannot account for this type of spectrum.  In a classical
Keplerian accretion disk (Shakura \& Sunyaev 1973), dissipated
mechanical energy is lost as local blackbody radiation, and there is a
simple relationship between the accretion rate $\dot M$ and the
luminosity $L$.  There is little freedom to adapt the model to more
complex spectra.

Another class of models is characterized by inefficient cooling and
high temperatures (e.g., Shapiro, Lightman \& Eardley 1976; Ichimaru
1977).  One particular model that of late has enjoyed considerable
attention is known as an Advection Dominated Accretion Flow, or ADAF
for short.  The epithet ADAF originated with the influential paper of
Narayan \& Yi (1994; hereafter NY94), which presented a
one-dimensional, time-stationary, self-similar solution (fixed
power-law dependencies in $r$ for dynamical variables).  All velocities
---radial, nonradial and thermal---are assumed to have the same
Keplerian $r$ dependence, $r^{-1/2}$.  Angular momentum transport is
effected by a Navier-Stokes viscosity along the lines of the Shakura \&
Sunyaev (1973) and Lynden-Bell \& Pringle (1974) $\alpha$ formalism.
The gas is hot, with the sound speed comparable to the orbital speed.
The internal energy is advected along with the mass into the central
black hole; very little is radiated.  An ADAF is more like a Bondi flow
than a Keplerian disk. The quasi-spherical nature of the flow is often
cited as a defining characteristic feature.  ADAFs are sufficiently
distinct from a Keplerian disk that a transition radius $r_t$ is
invoked to define the location separating the (outer) Keplerian flow
from the (inner) ADAF.

There are two dynamical features of the NY94 self-similar ADAF solution
that have attracted special attention.  The first is that
the specific energy flux of the gas (the Bernoulli parameter) is
everywhere positive, the second is that entropy increases inwards,
rendering the accretion convectively unstable by the Schwarzschild
criterion.  We review these in turn.

In a nonradiative accretion flow a positive Bernoulli parameter is a
consequence of the fact that the stress transports energy,
as well as angular momentum, outwards.  In a standard thin disk
solution, this transported energy is radiated on its outward journey,
so that the local radiated flux (outside of the inner boundary layer)
is always greater by a factor of 3 than the energy released locally.
By contrast, in a nonradiative accretion flow the transported energy is
retained, locally raising the specific energy.  If the flow originates
at large radius as a marginally bound fluid, the Bernoulli parameter
will accordingly be everywhere positive.  NY94 noted that a positive
Bernoulli parameter might make outflows possible, but since an outflow
is very distinct from their ADAF solution, the significance of this
observation is unclear.  Blandford \& Begelman (1999) argued that this
property is likely to change fundamentally the dynamical flow
properties of the nonradiative solution, and proposed that not only
were outflows possible, but that the {\em bulk} of the gas and energy
would be carried off by a wind.  They refer to this concept as an
Adiabatic Inflow-Outflow Solution, or ADIOS.

More recently, attention has focused upon the increasing-inward entropy
profile that is a straightforward consequence of viscous dissipation,
and is a property of any nonadiabatically heated, nonradiative
inflow.   Narayan \& Yi (1995) argued that radial convection would
inevitably arise, and serve as a source of {\em outward} angular
momentum transport---either as the self-consistent source of $\alpha$,
or at a minimum, as a supplement to it.  Narayan, Igumenshchev, \&
Abramowicz (2000) later explored the consequences of convection moving
angular momentum either outward or inward.  In one scheme, inward
angular momentum transport by convection is envisioned to be capable of
balancing the outward $\alpha$ transport, creating zero (net) mass and
angular momentum flux.  Quataert \& Gruzinov (2000) argued that
convection would produce a state of near marginal {\em convective}
stability rather like stellar convective zones, described by the
classical H\o iland criteria.  Quataert \& Gruzinov referred to this new
variation as a Convective Dominated Accretion Flow (CDAF), though
formally there is no mass accretion at all!

We regard the defining property of a specific flow to
be the fate of the accreting mass and energy.  In the Shakura-Sunyaev
thin disk model, liberated mechanical energy is both transported
outwards and radiated locally.  In an ADAF, this liberated energy is
advected with the flow
through the event horizon of the black hole.  In a CDAF, the accretion
flow is effectively stifled with no significant mass inflow or
outflow.  What little energy is extracted from the inner region of the
flow is simply transported outward by convection to large
radius.  In an ADIOS the net accretion is very small because systematic
outflows remove the bulk of the mass and energy.  The existence of what
seem to be under-luminous X ray sources lends credence to the existence
of nonradiative flows.  The question is, are they
best described as ADAFs, CDAFs, ADIOS, or something else?

While flow energetic properties are likely to be complex and (for the
moment) highly model-dependent, gross dynamical features---inflow
versus outflow, rotational versus pressure support---should be more
robust.  In this work, we report the results of three-dimensional 
magnetohydrodynamic (MHD)
simulations that follow the accretion of rotating, nonradiating,
magnetized gas onto a black hole.
We find that the bulk of the mass in nonradiative flows is not
slowly rotating and quasispherical like an ADAF, but is much closer to
Keplerian:  rotationally supported and disk-like.  Most of the
liberated energy is  carried outward by streaming backflow and turbulent
transport.  The nature of the flow is entirely determined by MHD
turbulence, triggered by the magnetorotational instability (MRI; Balbus
\& Hawley 1991).

An outline of the paper is as follows.  In \S2 we discuss the numerical
model for the simulations.  In \S3 we present two such simulations, one
with resistive dissipation and one without.  In \S4 we summarize the
main features of our model, and discuss the implications of the
simulation for the accreting black hole in Sgr A*.  Our
conclusions are given in \S5.

\section{Numerical MHD Simulations: Background}

Rotating, non-radiative accretion flows (NRAFs) can be numerically
simulated.  A number of axisymmetric hydrodynamical simulations have
been performed (e.g., Igumenshchev \& Abramowicz 1999, 2000; Stone,
Pringle, \& Begelman 1999), that drive accretion by employing an
explicit kinematic viscosity, $\nu$.  The results depend strongly upon
both the specific recipe adopted for $\nu$, and its magnitude.  For
example, Igumenshchev \& Abramowicz (1999, 2000) found that large
viscosity flows accrete directly into the central hole.  High viscosity
accretion flows are therefore associated with the quasi-spherical ADAF
flows.   The low viscosity simulations
(Igumenshchev \& Abramowicz 1999, 2000; Stone et al. 1999), on the
other hand, have a much more difficult time reaching the central hole.
These flows have been interpreted as CDAFs.

MHD fluids, however, are fundamentally different from unmagnetized
fluids, and essential features cannot be modeled with a Navier-Stokes
viscosity formalism.  Accretion simulations must be MHD.

The first NRAF MHD simulations were done by Stone \& Pringle (2001;
hereafter SP).  These simulations were axisymmetric, and this is a
limitation.  First, the anti-dynamo theorem (e.g., Moffatt
1978) prevents the indefinite maintenance of a poloidal magnetic field
in the face of dissipation.  Indeed, toward the end of the SP
simulations, the turbulence begins to die down, persisting only in flow
close to the black hole.  Second, axisymmetric
simulations tend to over-emphasize the ``channel" mode (Hawley \&
Balbus 1992), which produces coherent streaming in the disk plane
rather than the more generic MHD turbulence.  Finally, the toroidal
field MRI cannot be simulated in axisymmetry.  Consequently, a fully
self-consistent accretion simulation requires three dimensional MHD.

\subsection{Equations}

The simulations described in this paper evolve the three-dimensional
equations of MHD: the continuity equation, the equation of motion, an
internal energy equation, and the induction equation:
\begin{equation}\label{mass}
{\partial\rho\over \partial t} + \nabla\cdot (\rho {\bf v}) =  0,
\end{equation}
\begin{equation}\label{mom}
\rho {\partial{\bf v} \over \partial t}
+ (\rho {\bf v}\cdot\nabla){\bf v} = -\nabla\left(
P + {\mathcal Q} +{B^2\over 8 \pi} \right)-\rho \nabla \Phi +
\left( {{\bf B}\over 4\pi}\cdot \nabla\right){\bf B},
\end{equation}
\begin{equation}\label{ene}
{\partial\rho\epsilon\over \partial t} + \nabla\cdot (\rho\epsilon
{\bf v}) = -(P+{\mathcal Q}) \nabla \cdot {\bf v} + \eta J^2,
\end{equation}
\begin{equation} \label{ind}
{\partial{\bf B}\over \partial t} =
\nabla\times\left( {\bf v} \times {\bf B} -\eta_i {\bf J} \right),
\end{equation}
where $\rho$ is the mass density, $\epsilon$ is the specific internal
energy, ${\bf v}$ is the fluid velocity, $P$ is the pressure, $\Phi$
is the gravitational potential, ${\bf B}$ is the magnetic field
vector, ${\bf J}$ is the current, 
${\mathcal Q}$ is an explicit artificial viscosity of
the form described by Stone \& Norman (1992a), and $\eta$ is an
anomalous resistivity of the form used by SP, namely
\begin{equation}\label{artres}
\eta_i = C_{res}(\Delta x_i)^2 |J_i|/\sqrt{\rho}.
\end{equation}
The constant $C_{res}$ needs to be large enough to spread a current
sheet out over a few zones, but not so large as to turn the overall
flow into a resistive one.
We use a resistivity constant of $C_{res} = 0.1$, as did SP.

The form of $\Phi$ is the pseudo-Newtonian gravitational potential 
introduced by Paczy\'nski \& Wiita (1980), 
\begin{equation}\label{poten}
\Phi = - GM/(r-r_g),
\end{equation}
where $M$ is the mass of the central black hole, and $r_g$ is the
gravitational radius, equivalent to the Schwarzschild
radius in general relativity.
With this potential, the angular momentum of a circular orbit is
\begin{equation}\label{lkep}
l_{Kep} = (GMr)^{1/2} r/(r-r_g),
\end{equation}
and  the binding energy is
\begin{equation}\label{bindinge}
e= -\left({GM\over2rc^2} \right)\left( {[r-2r_g] r \over [r-r_g]^2}\right)
c^2.
\end{equation}
The pseudo-Newtonian potential mimics the dynamically important
marginally stable circular orbit 
of the full Schwarzschild metric 
(defined by $dl_{Kep}/dr = 0$) 
at $r=r_{ms}=3r_g$.

The equation of state is $P=\rho\epsilon (\gamma -1)$, with
$\gamma=5/3$.  Radiation transport and losses are by assumption
dynamically unimportant in an NRAF, and are omitted.  There is no
explicit shear viscosity; angular momentum is transported by Maxwell
and Reynolds stresses arising from magnetic and velocity correlations
in the MRI-induced turbulence.  Similarly, the gas is not heated
directly by an $\alpha$-viscosity; nonadiabatic heating comes from the
artificial viscosity ${\mathcal Q}$ and the resistivity.   Both
allow the entropy of the gas to increase.

We evolve the equations using time-explicit Eulerian finite
differencing with the ZEUS algorithms (Stone and Norman 1992a,b; Hawley
\& Stone 1995).

\subsection{Physical Units and Code Units}

The results presented below are generally presented in terms of scale-free
code units.  Time is measured in orbital periods at the location of the
pressure maximum of the initial torus, $R=100r_g$.  (Here, $R$ is the
cylindrical radius.)  This is 286 orbital periods at the marginally
stable orbit, $r_{ms}$.

It is often convenient to have astrophysical scales associated with
these values.
Following Paczy\'nski \& Wiita (1980), 
we equate the gravitational radius $r_g =1$ with 
the Schwarzschild radius,
\begin{equation}
r_g = 3 \times 10^{5} \left( M/M_\odot\right) \ {\rm cm}
\end{equation}
where $M/M_\odot$ is the black hole mass.  We also set $GM=1$, and
the speed of light is $c=(2GM/r_g)^{1/2}$, or in code units $\sqrt 2$.
The orbital time at $100r_g$ is 
$$
\simeq 2\pi \times 10^3 \sqrt{2} r_g/c.
$$
This is defined to be $2\pi\times 10^3$ code units of time, so that 
one code time unit is
$\sqrt{2}r_g/c\simeq 14 M/M_\odot$~$\mu$s.  For a $2.6\times 10^6
M_\odot$ hole, one orbit at $R=100r_g$ is $2.3\times 10^5$~s; the
entire simulation covers about 18 days in the life of the accretion
flow.

In the absence of self-gravity and radiation, there is no density scale.
The physical number density can be either specified outright, 
set by the mass of the initial torus, or assigned a value by
specifying the accretion rate.
The accretion rate 
can be scaled by the usual Eddington luminosity
\begin{equation}
L_{Edd} = 1.3\times 10^{38} M/M_\odot \ {\rm erg\ s^{-1}},
\end{equation}
and the Eddington mass accretion rate of
\begin{equation}
\dot M_{Edd} = L_{Edd}/c^2 =  1.4\times 10^{17} M/M_\odot \ {\rm gm\ s^{-1}}.
\end{equation}

Since a nonradiating rotationally-supported gas is approximately virial
in a black hole potential, the gas temperature will generally be of
order the binding energy (\ref{bindinge}), a significant fraction of
the rest mass energy (recall that $m_p c^2/k \simeq 10^{13}$K).  In the
pseudo-Newtonian potential, the binding energy of the marginally stable
orbit is $0.0625c^2$.  In our single fluid calculation there is only
one temperature, $T$, and no distinction is made between ion and
electron temperatures.  A gas simulation such as this provides no
constraints on the electron-ion interaction.  However, if the ions
provide all the dynamical pressure, it would be straightforward to
consider the consequences of a two temperature plasma where the
electron temperature $T_e$ is some fraction $\delta$ of the ion
temperature $T_i$.

\section{Three Dimensional NRAF Simulations}

A previous paper (Hawley, Balbus, \& Stone 2001) presents a prototype
MHD NRAF simulation.  Here we investigate that simulation (designated F1)
in detail, and extend it to later times.  We also compare F1 with a
simulation that includes resistive dissipation (simulation F1r).  
The motivation for this is that in CDAF models 
dissipative heating in the central regions is argued to be
responsible for global flow structure (Narayan et al. 2000).

We begin with some technical details.  These simulations use
$128\times32\times 128$ grid zones in cylindrical coordinates
$(R,\phi,z)$.  The radial grid extends from $R=1.5r_g$ to $R=220r_g$.  There
are 36 equally-spaced zones inside $R=15r_g$, and 92 zones increasing
logarithmically in size outside this point.  There are 50 $z$ zones
equally spaced between $-10$ and $10$ with the remainder of the zones
logarithmically stretched to the $z$ boundaries at $\pm 60$.  The
azimuthal domain is uniformly gridded over $\pi/2$ in angle.  The
radial and vertical boundary conditions are simple zero-gradient
outflow conditions; no flow into the computational domain is
permitted.  The $\phi$ boundary is periodic.  The magnetic field
boundary condition requires the transverse components of the field to
be zero outside the computational domain, while the perpendicular
component satisfies the divergence-free constraint.

The initial condition is a constant specific angular momentum ($l$)
torus with a pressure maximum at $R=100r_g$, an inner edge at
$R=75r_g$, and an outer edge at $R=153r_g$.  The initial magnetic field
consists of poloidal loops lying along isodensity contours.  It has a
volume-averaged $\beta$ value $(=P_{gas}/P_{mag}$) of 200.  The average
specific energy of the magnetofluid in the torus is $-2.3\times
10^{-3}c^2$, i.e., it is bound.  Simulation F1 is run for 6.32 orbits
at the initial pressure maximum, or 1,807 orbits at $r_{ms}=3r_g$.
Simulation F1r begins at a time equal to 
3.5 pressure maximum orbits in F1, after a global flow has
been established.  It continues to a time of 6.38 orbits.

\begin{figure}
\plotone{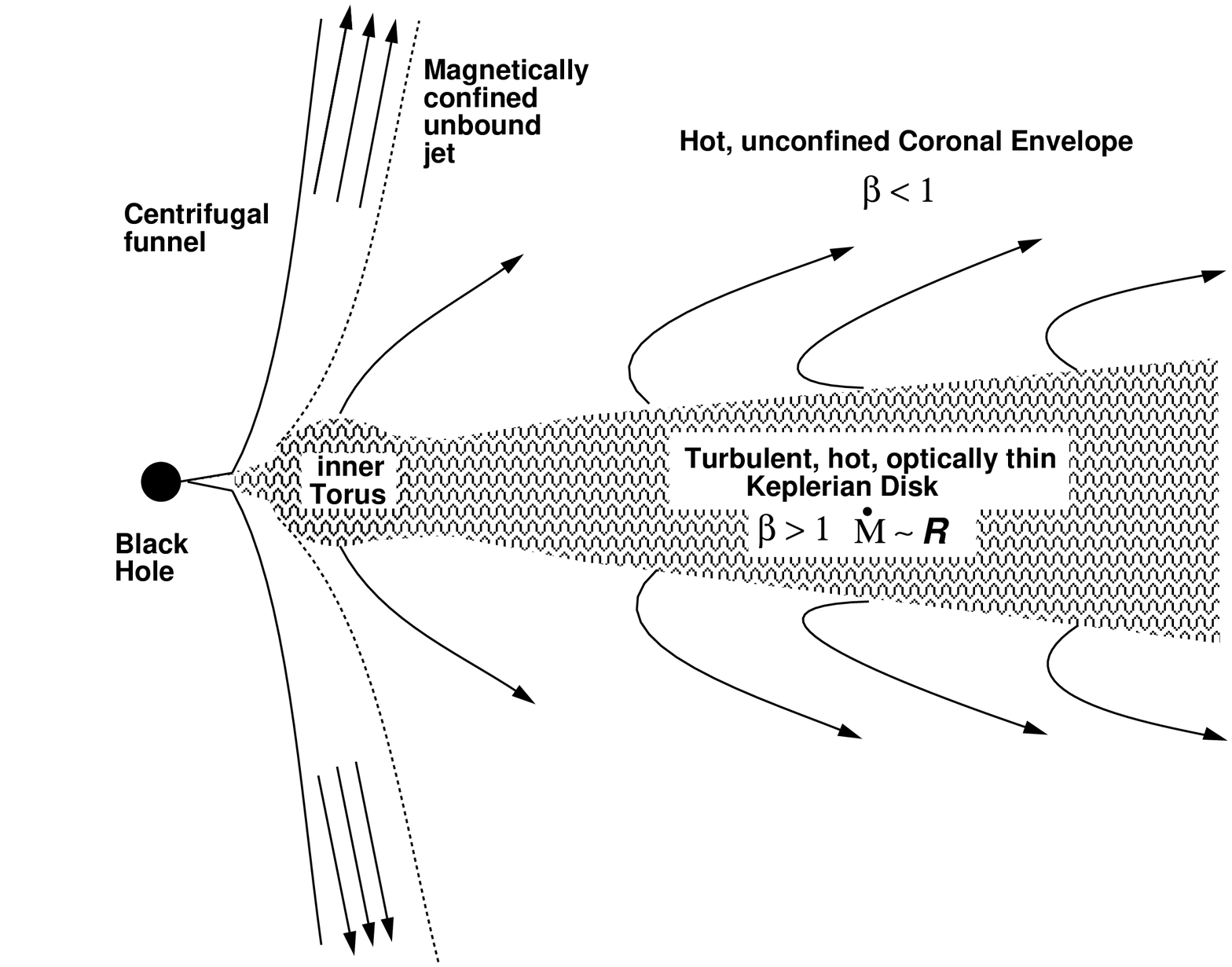}
\caption{A schematic diagram of the NRAF showing its major dynamical
structures.  An (MRI) turbulent, nearly Keplerian hot disk is
surrounded by a dynamic low density envelope.  Near the marginally
stable orbit, the accretion flow thickens into a small inner torus as
the pressure rises, but is still primarily supported by rotation. A
centrifugally-evacuated funnel lies along the axis and is surrounded
by an unbound outflowing jet confined by magnetic pressure in the
corona.  }
\label{schematic} 
\end{figure}

Our simulations, with or without resistive heating, show three
principal flow components:  (1) a hot, but rotationally-supported, disk
extending down to the marginally stable orbit; (2) an extended, low
density coronal backflow enveloping the disk; (3) a distinctive,
jet-like flow near the hole, that emerges unambiguously in momentum
plots.  These features are illustrated schematically in
Figure~\ref{schematic}.  We believe that these structures are real
and robust, and that they are fundamental generic properties of NRAFs.
Physical arguments underlying the
origin of NRAF structure are presented in \S4.1.

\subsection{Simulation F1: Ideal MHD}

\begin{figure}
\plotone{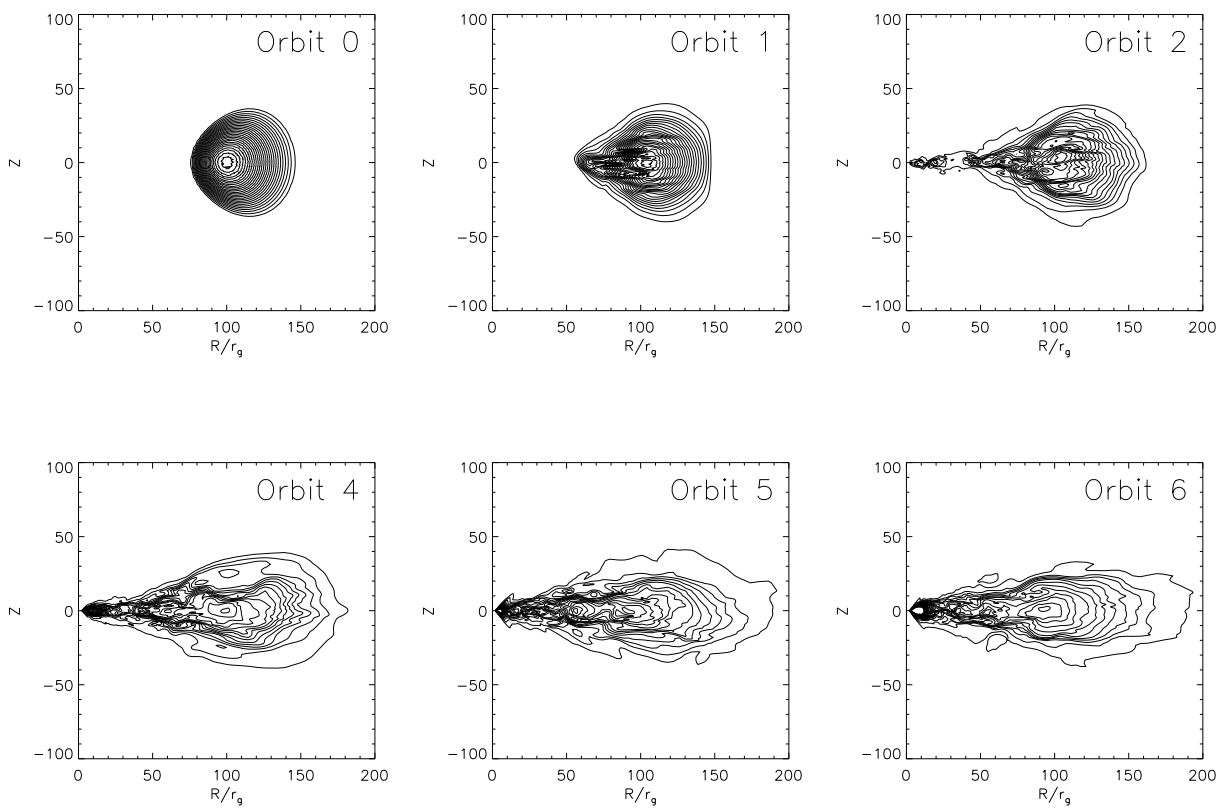}
\caption{Density contours at initial pressure-maximum 
orbits 0, 1, 2, 4, 5, and 6.  There are
30 contours, linearly spaced between 0 and 1.2.  (The maximum density
in the initial torus is defined to be 1.)  During orbits 1 and 2 the
linear modes of the MRI can be clearly seen.  
Beginning with orbit 4, a thinner disk structure is established; this disk
stretches out radially with advancing time.  By orbit 6 the hot inner
torus can be seen inside of $R=10r_g$.}
\label{densevol}
\end{figure}

The evolution begins as a thick torus, with $H/R = 0.2$.
The complete evolution is represented in Figure~\ref{densevol} 
as a series of contour plots in azimuthally-averaged
density.  During the first orbit, the field in the torus is amplified
both by the MRI and by shear, and the flow evolves rapidly.  The MRI acts
most effectively near the equatorial plane, where the field is
predominantly vertical.  Here the long-wavelength, nearly axisymmetric
modes of the MRI grow rapidly.  Significant accretion begins shortly
after one orbit, when the magnetic energy in the torus has increased to
$\beta \approx 2$--10.

Between orbit 1 and 2, low-$m$ (azimuthal wavenumber) spiral arms of
gas accrete from the inner edge of the torus, forming a vertically thin
and very nonuniform ``daughter disk.'' Strong magnetic fields surround
the gas.  Due to the initial field topology,  a current sheet forms
near the equator, which proves unstable to vertical oscillations.  At
later times, as more gas accretes from the initial torus, the disk
fills out and thickens.  Inside of $R=100r_g$ the NRAF forms a modestly
thick, nearly Keplerian disk.  Low density material is stripped off this
evolving disk to create a backflow.  Inside of $R=100r_g$, the net mass
flux is inward; for $R>100r_g$ there is net outflow.

By orbit 2, the mass accretion rate into the central black hole has
approached its long-term average value, which is only few percent of
the rate at which the torus feeds mass into the inner region.  There
are two requirements for the gas to accrete into the central black
hole, and neither is easy to meet: (1) the angular momentum of the gas
must be reduced to a value close to that of the marginally stable
circular orbit; (2) the gas also must be relatively cool, or pressure
and centrifugal forces will drive the gas back out.  In this simulation
this is much in evidence:  accreting gas must pass through a narrow gap
in the equatorial ``waist'' of the centrifugal barrier.  When the gas
is hot and geometrically thick, this is most difficult.  Instead of
accreting, much of the gas splashes off the centrifugal barrier near
the hole, either contributing to the formation of a hot, intermittent
torus inside of $R=10r_g$, or creating a coronal backflow that adds to
a growing low density envelope around the Keplerian core disk.

\begin{figure}
\plottwo{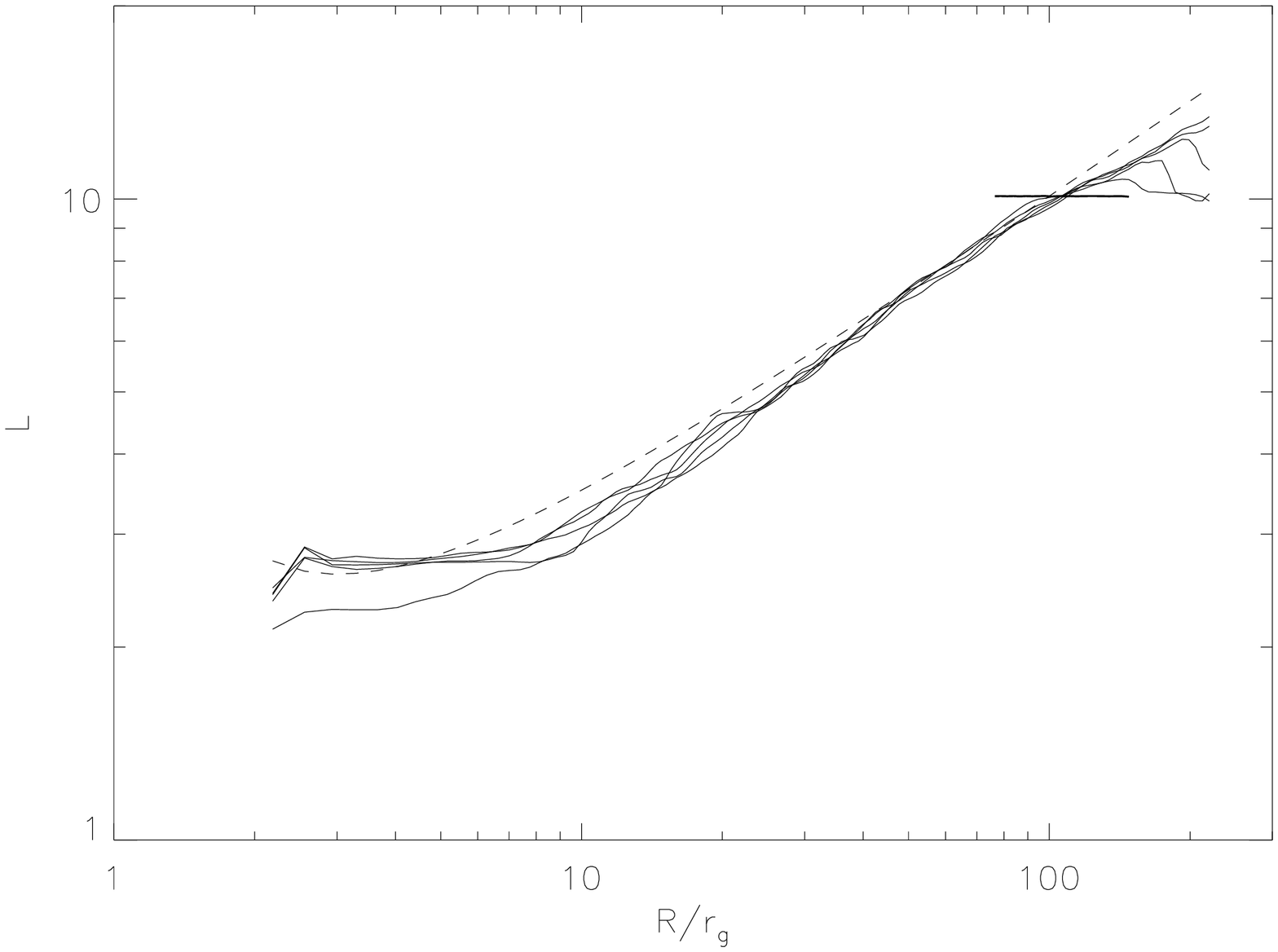}{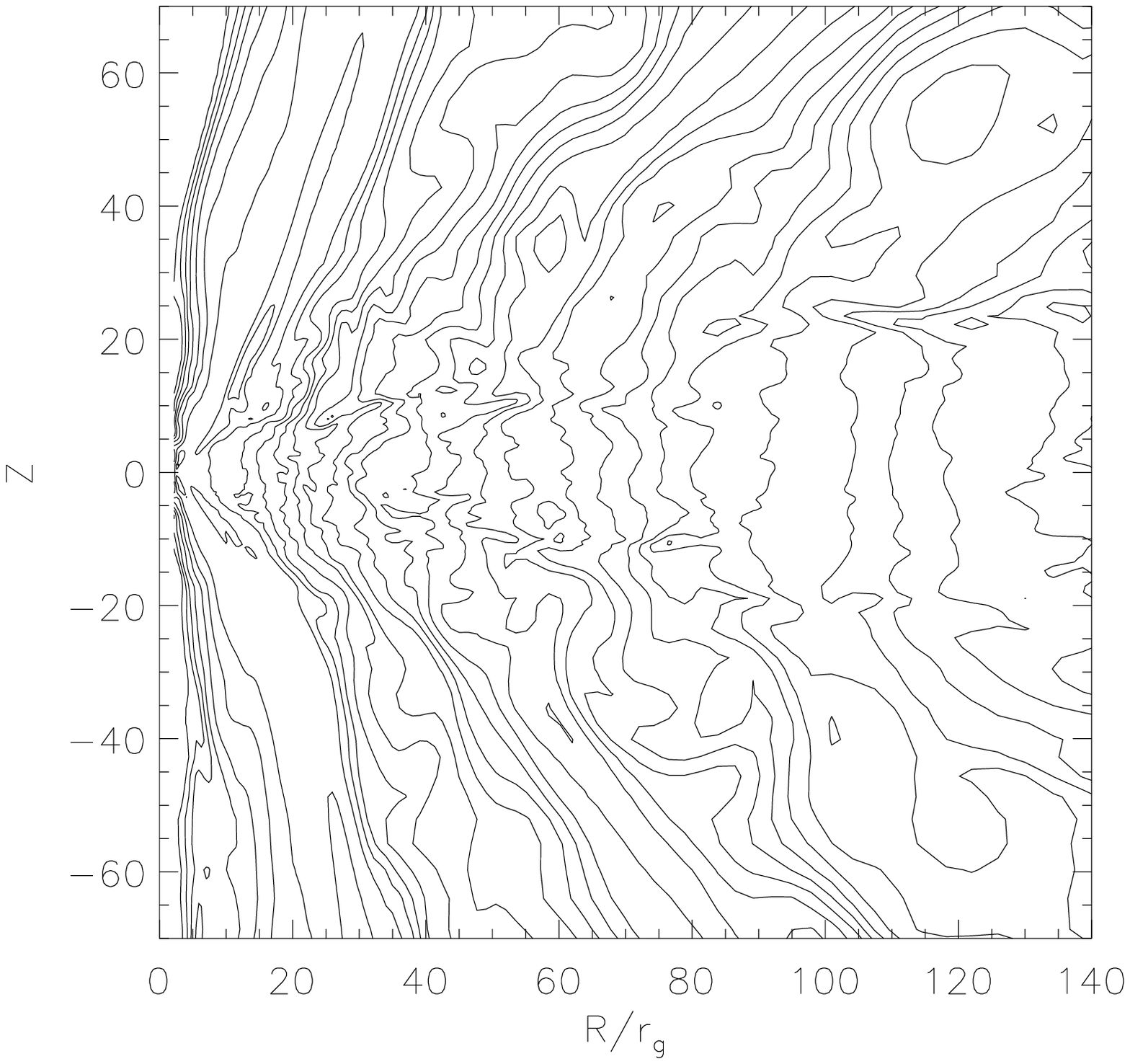}
\caption{Evolution of the specific angular momentum.  (a) The
mass-weighted, vertical- and azimuthally-averaged specific angular
momentum $\langle l \rangle$ is plotted as a function of radius at
each orbital time at the pressure maximum.  
For reference, the dashed line depicts the value $l_{Kep}$
corresponding to a circular orbit.
At $t=0$ $\langle l \rangle$ is a constant 
($=10.1$).  After only one orbit of time the flow has 
established a nearly Keplerian distribution throughout the radial
domain, which it maintains throughout the simulation.
(b) Contours of azimuthally-averaged specific angular
momentum at the end of the simulation.  There are 40 linearly
spaced contours between $l=0.$ and $l=14.$}
\label{angmom}
\end{figure}

As time advances, the equatorial inflow follows a nearly
Keplerian angular momentum distribution, $l_{Kep} \propto R^{1/2}$ (see
fig.~\ref{angmom}).  The value of $l$ actually remains everywhere
slightly sub-Keplerian, except at the innermost radii where it is
slightly super-Keplerian (in the hot inner torus).  At the end of the
evolution, $l$ is about $\lta$ 5\% below $l_{Kep}$ between $R\sim
30r_g$ and $100r_g$, but for $R\sim 10r_g$ $l$ drops to as much as 15\%
below the Keplerian value.

Between orbits 4 and 5, the accretion flow matures and becomes
reasonably steady, forming a moderately thick disk surrounded by a low
density, highly magnetized atmosphere.  Between orbit 5 and the end of
the run (6.3 orbits), the properties in most of the disk do not change
by large amounts.  The exception to this general trend is the inner
torus ($R \le 10r_g$), which sporadically depletes and reforms.

\begin{figure}
\plotone{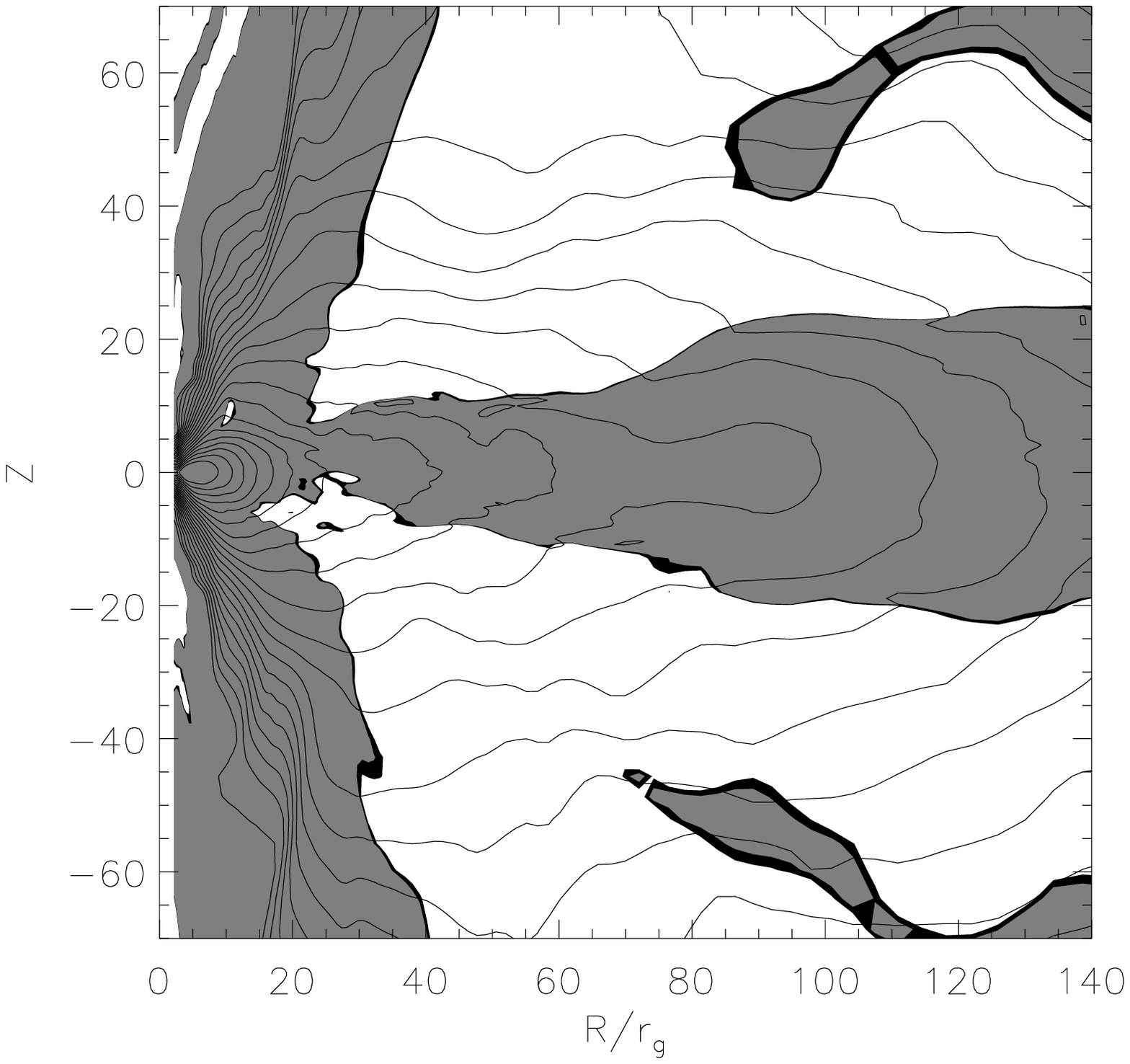}
\caption{Contours of the logarithm of total pressure, magnetic plus gas, 
at the end of the F1 simulation.
The shaded regions overlaid on the contours show where gas pressure
exceeds magnetic, i.e., $\beta \ge 1$. 
The bulk of the coronal envelope is magnetically dominated.
Gas pressure dominates in the disk, the hot inner torus, and in the
jet along the funnel wall.}
\label{tpressure}
\end{figure}

Near the equator, thermal pressure dominates magnetic, i.e., $\beta >
1$, whereas the surrounding region is strongly magnetized, $\beta < 1$
(fig.~\ref{tpressure}).  The thermal pressure exponential scale height
is $H\approx 7$--10$r_g$, decreasing rapidly inside $R=10r_g$.  The
total pressure (thermal plus magnetic) is much smoother, and has an
exponential scale height $H\approx 20r_g$ at $R=100r_g$; $H$ decreases
slowly inward.  The gas density, vertically averaged over one gas
pressure scale height, is nearly constant from $R=50r_g$ down to
$R=10r_g$, where it increases rapidly to a peak at $R=5.5r_g$.  Inside
of $R=10r_g$ the disk resembles a thickened torus.

The disk and coronal regions can also be distinguished in Figure
\ref{angmom}b which shows contours of specific angular momentum.
Within the disk the gas is approximately rotating on cylinders,
characteristic of a nearly barotropic equation of state.   In the
corona, surfaces of constant specific angular momentum are swept back,
consistent with conserved $l$ along outflowing streamlines.

\begin{figure}
\plotone{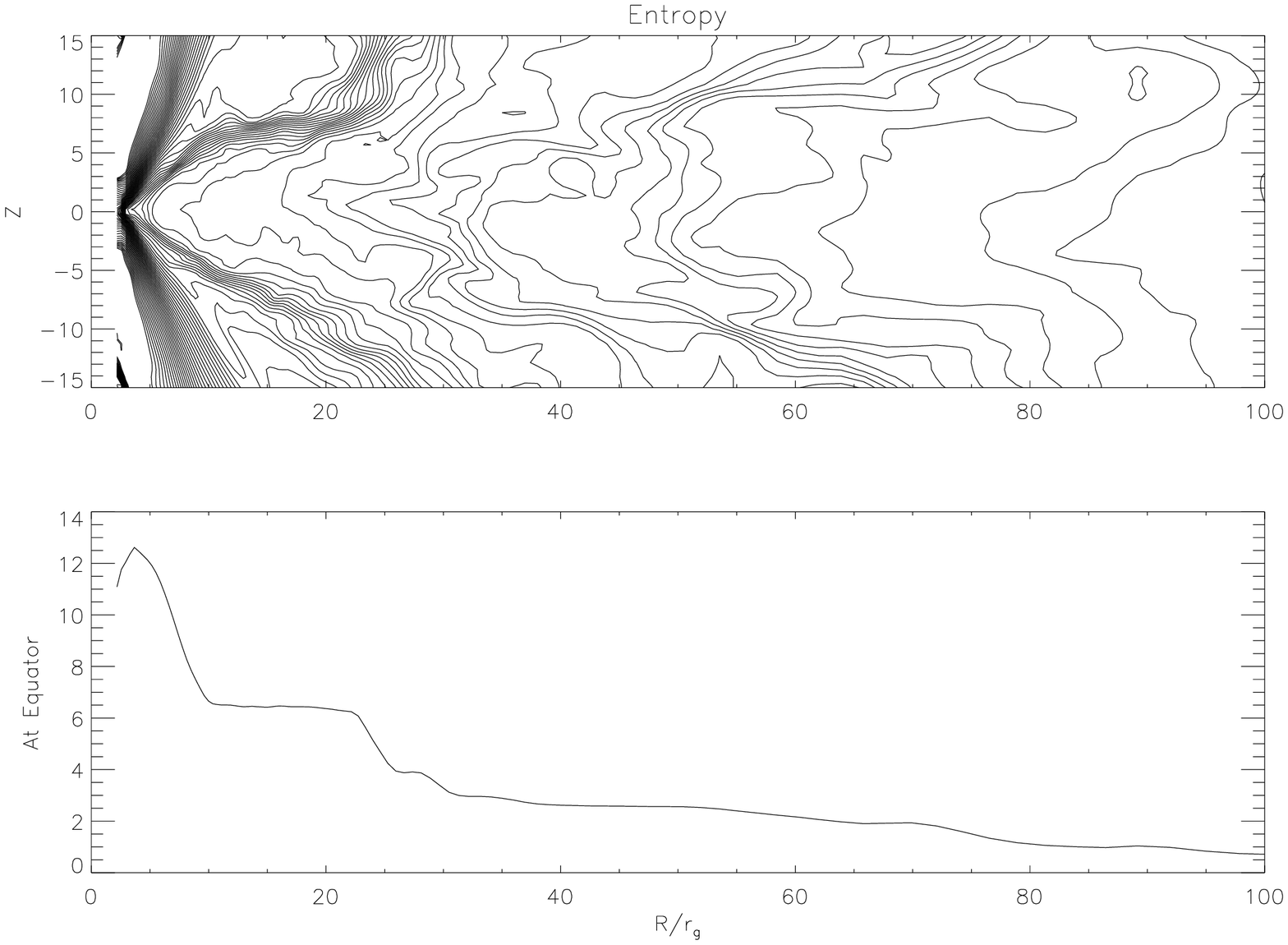}
\caption{Contours of azimuthally-averaged entropy 
(top), and along the equator (bottom) at the end of the F1
simulation.
The values are normalized to the initial entropy of the torus, $S(t)-S_i$.}
\label{entropy}
\end{figure}

Figure~\ref{entropy} shows contours of entropy, $S \propto \ln
(P\rho^{-5/3})$, and the entropy along the equatorial plane at the end
of the F1 simulation.  Note that because of nonadiabatic heating by the
artificial viscosity, the entropy increases inward even in the absence
of resistive heating.

Magnetic fields in the disk produce a significant Maxwell stress which
results in turbulent angular momentum transport.  Between orbit 4 and
the end of the simulation, the ratio of the vertical- and
$\phi$-averaged Maxwell stress to the gas pressure (a form of the
$\alpha$ stress parameter) ranges from $\sim 0.05$ to as much as 0.2
inside of $R=100r_g$.  This is significantly higher than the values
found in small local disk simulations with zero net field 
(e.g., Hawley, Gammie, \& Balbus 1996).  It is this
vigorous outward angular momentum transport that is responsible for
building up a Keplerian profile:  an initially shallower angular
momentum distribution, as is typical in a thick torus equilibrium,
exports more and more of its angular momentum to the exterior, until
much of the radial pressure support is eliminated.  The result is a
Keplerian profile.

\begin{figure}
\plotone{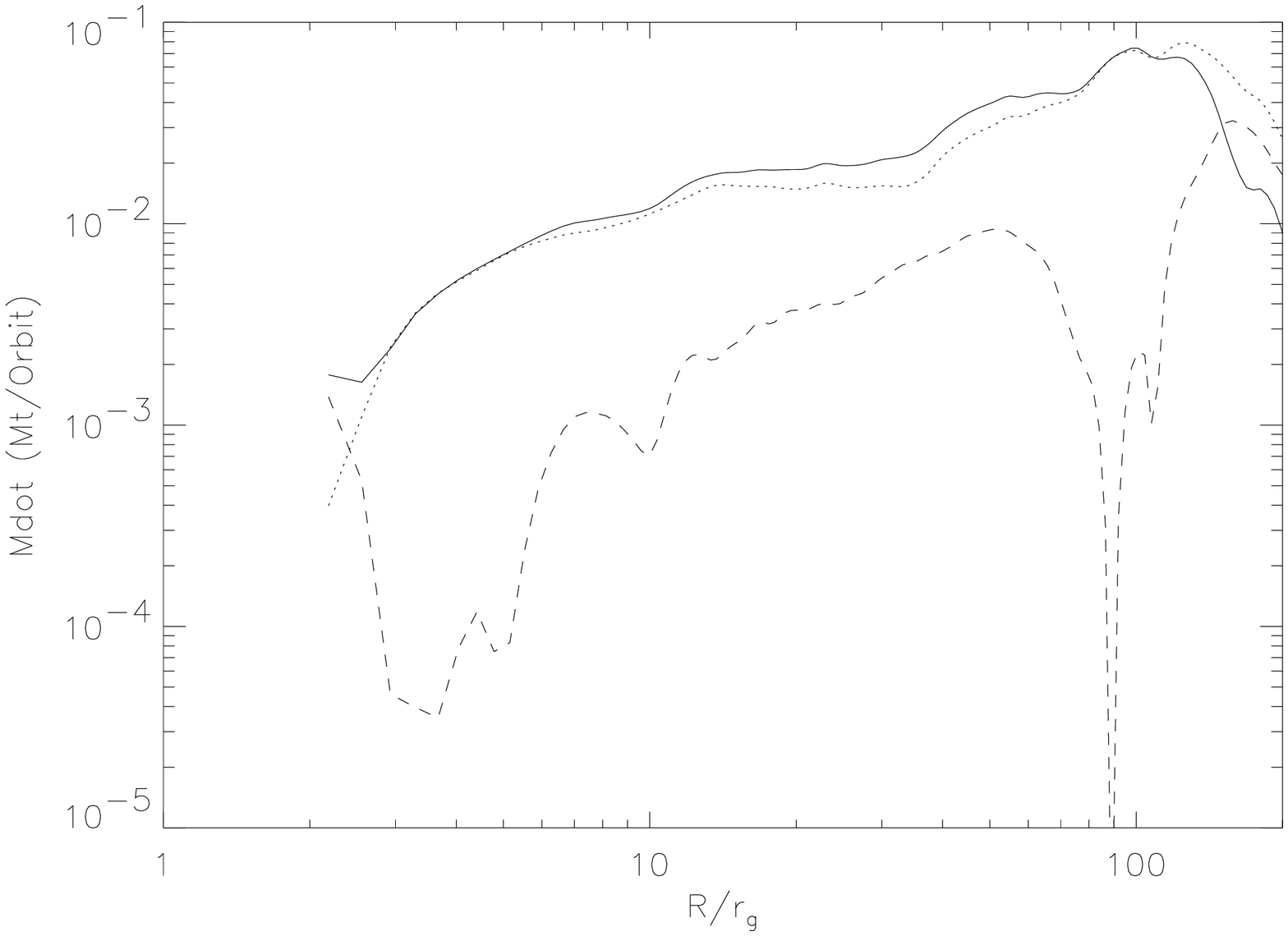}
\caption{Accretion rate $\dot M$ averaged over the last orbit of
time.  The solid line is the mass inflow rate, the dotted line the
mass outflow rate, and the dashed line the difference, i.e., the net
accretion rate.  Between $R=10r_g$ and $R=60r_g$ the net accretion rate is
nearly proportional to $R$.}
\label{mdot}
\end{figure}

The mass inflow ($v_R < 0$) and outflow ($v_R > 0$) rates 
through each cylindrical radius are computed as a function of time,
\begin{equation}\label{dotm}
\dot M = \int \int R \rho  v_R dz d\phi .  
\end{equation}
The instantaneous local inflow and outflow rates are primarily due to
the mass flux produced by short term $\rho$ and $v_R$ fluctuations, and
they are nearly equal.  This is not a matter of systematic inward mass
flow being cancelled by a systematic outward mass flow.  Rather, this
is intrinsic to the nature of turbulent flow.  The long term {\em net}
accretion rate, $\dot M$, is always smaller than its rms value.  An
average over time for the last orbit of the simulation shows that the
inflow exceeds the outflow for $R<90r_g$ (fig.~\ref{mdot}).  The ratio
of inflow to outflow is about 1.5 between $R=30r_g$ and $90r_g$,
dropping toward unity as one moves inward until $r_{ms}$ is reached, at
which point the outflow rate plummets to zero.  The time-averaged net
accretion rate $\dot M$ is roughly proportional to $R$ between
$R=10r_g$ and $60r_g$, i.e., it is {\it not constant.}  The net $\dot
M$ can vary with $R$ if either the flow is not steady, e.g., the mass
in the disk is building up with time, or if there is flow out of the
grid at the $z$ boundaries.  If we consider the computational volume
bounded by $R=90r_g$ on the outside over the last orbit of time, we
find that the total mass in this volume declines and that over half of
the net loss is through outflow along the $z$ boundaries.  We conclude
that much of the decline of $\dot M$ with radius can be accounted for
by this outflow.

The total mass on the grid has decreased by 8.7\% by the end of the
simulation.  Roughly 51\% of this leaves through the outer radial
boundary, 44\% through the upper and lower $z$ boundaries, and the
remaining 5\% is accreted into the black hole.  Integrating the mass
flux at $R=75r_g$ over time shows that a net 14\% of the initial torus
mass has moved inward.  Only about 3\% of the material accreting
inward from the initial torus ends up in the central black hole.

\begin{figure}
\plotone{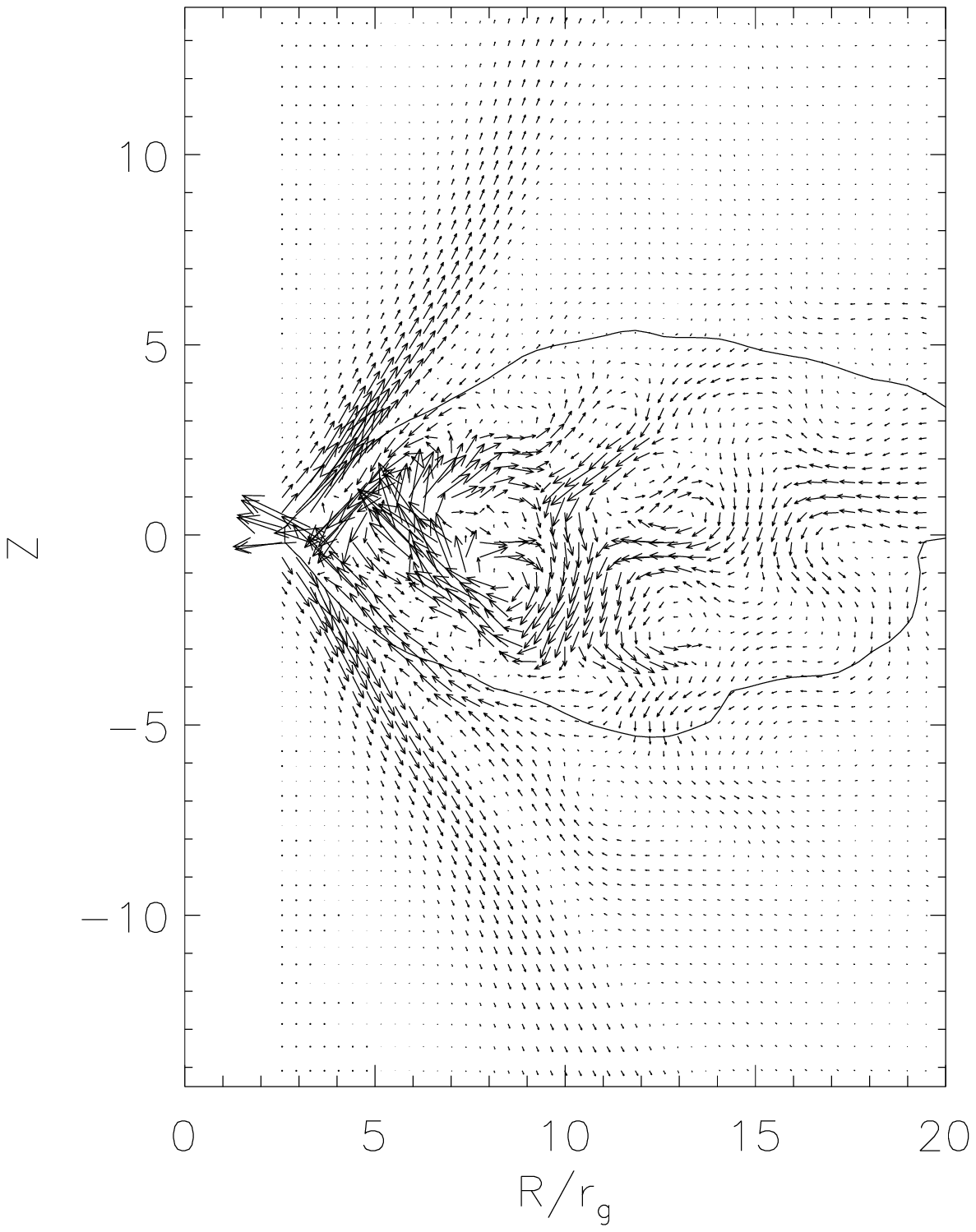}
\caption{Azimuthally-averaged momentum vectors in the inner region, $R
< 20r_g$, at the end time of simulation F1.  The overlaid contour is density, 
showing the shape of the inner torus.  The outflow along the funnel 
wall is clearly evident.  This
outflow is confined by magnetic pressure in the corona.}
\label{momentum}
\end{figure}

Much of the flow through the outer radial boundary is a consequence of
the increasing specific angular momentum in the outer part of the
torus.  The flow through the $z$ boundaries, however, is driven from
the disk by thermal and magnetic pressure.  Perhaps the most
interesting part of this outflow is an unbound, high temperature hollow
conical outflow, confined to the axis region by surrounding magnetic
pressure (see fig.~\ref{tpressure} and fig.~\ref{momentum}).  
By orbit 5, 60\% of the outflow
through the $z$ boundaries is in this conical wind.  The remainder of
the $z$ outflow is bound, although its specific energy is {\it greater} 
than that of the initial torus.  Whether or not this more widespread
outflow can be an unbound wind may therefore be a question of initial
conditions: if the flow had begun as a marginally bound gas, the outflow
would be unbound.

The initial average specific energy for the gas is $-2.3\times
10^{-3}c^2$.  After 5 orbits this has decreased to $-2.5\times
10^{-3}c^2$ as the gas remaining on the grid becomes more bound.  The
total thermal, magnetic, poloidal kinetic, and orbital energies of this
gas all increase with time.  The magnetic energy has increased the
most, receiving 42\% of the total net energy increase; 95\% of this is
in toroidal field.  The thermal, orbital and kinetic energies comprise
25\%, 21\%, and 12\% of the increase respectively.

\subsection{Simulation F1r: Resistive Heating}

In this calculation, simulation F1 is rerun beginning at orbit 3.5 with
the addition of an explicit resistivity,
\begin{equation}
\eta_i = 0.1 (\Delta x_i)^2 |J_i|/\sqrt{\rho}.
\end{equation} 
We refer to the resistive simulation as F1r.  The resistivity enhances
smallscale reconnection and returns the magnetic energy losses
as bulk heating.  A study of the effect of resistive heating is
motivated in part by claims that such heating will lead to convective
instabilities.  Resistive heating and turbulent dissipation are
physical processes hydrodynamical codes try to mimic using a
Navier-Stokes viscosity.  In hydrodynamic simulations, this $\alpha$
viscosity is argued to be the origin of convective heating and additional
angular momentum transport (Narayan et al. 2000).

Full MHD simulations reveal almost no difference between the resistive
F1r model and the nonresistive F1 model described above.  There is no
evidence of convective transport due to resistive heating.  This result
was already obtained by the 2D MHD simulations of SP, who compared runs
with and without resistivity.  Although the simulation outcomes will
always depend somewhat on the specific implementation of the
resistivity, grid resolution and the like, there are good reasons to
believe that thermal convection {\it per se} is generally of little
dynamical importance in nonradiating accretion flows (Balbus \& Hawley
2002).  They are dominated by MHD turbulence.

There are some minor differences between F1 and F1r that show up in the
details.  Immediately after switching on the resistivity, for example,
the magnetic field energy in F1r declines, and there is a corresponding
increase in the thermal energy.  The changes are small.  Integrated
over the entire computational domain, the thermal energy increases by a
maximum of 5\% over run F1 at 4.5 orbits.  As the simulation proceeds,
however, the total thermal energy in F1r actually lags  behind that of
F1.  By orbit 6, F1 has 17\% more thermal energy than F1r.  Model F1
builds up a hotter inner torus, in part because of a larger accretion
rate into the inner region.  Heating via the artificial viscosity in
shocks is more important than heating by the explicit resistivity.  The
shocks that occur in the simulation are not large-scale, coherent, or
time-independent.  The nonadiabatic heating appears to arise mainly in
small-scale, highly time-dependent dissipation of the turbulence.

\begin{figure}
\plotone{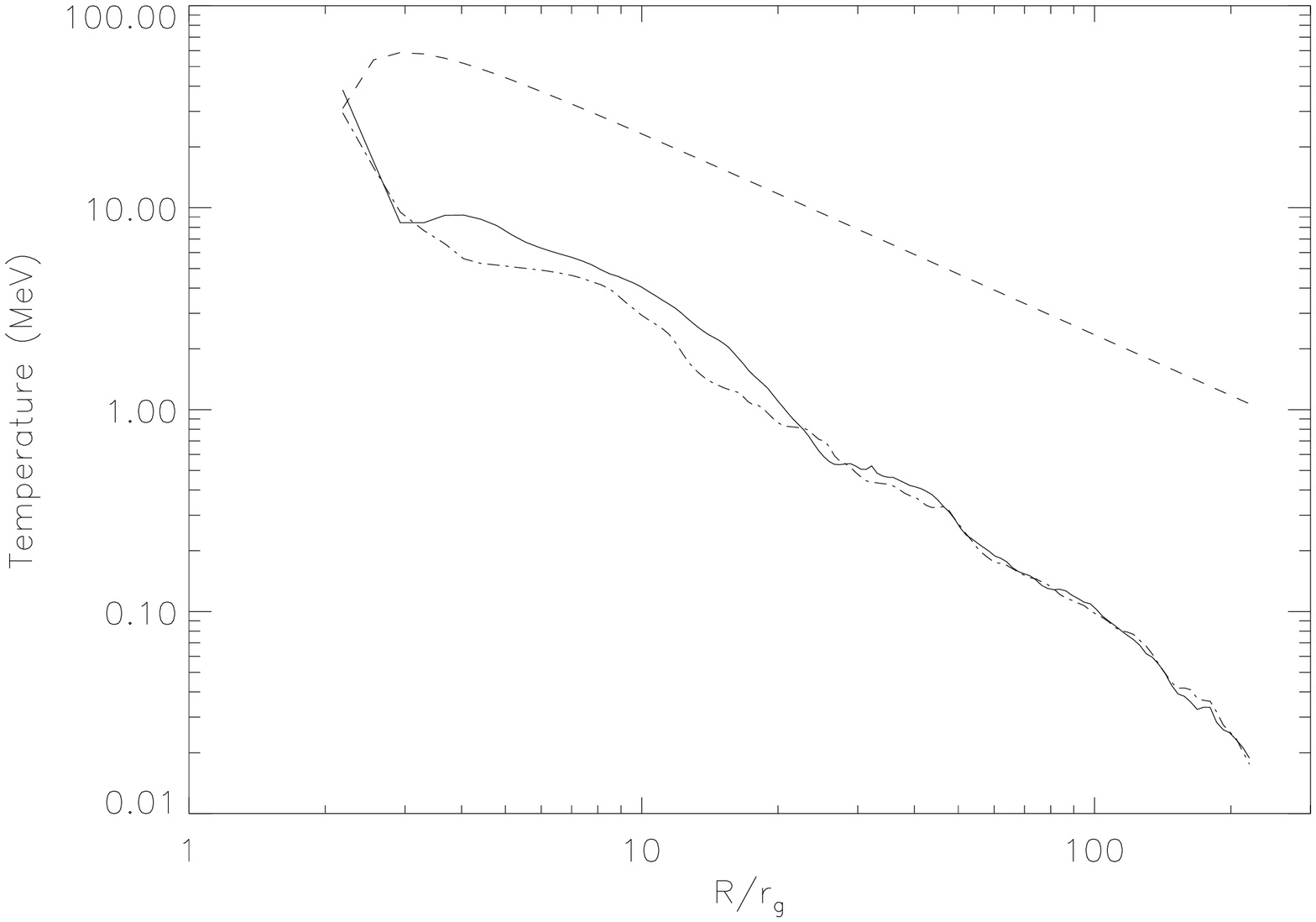}
\caption{Temperature at the equator as a function of radius at the
endpoints of simulation F1 (solid line) and F1r (dot-dash line).
For reference the binding energy is also plotted
(dashed line).  The maximum binding energy of $0.0625 m_p c^2$
occurs at $R=r_{ms} =3r_g$. }
\label{temperature}
\end{figure}

Although resistive dissipation increases the overall thermal energy, it
must do so at the expense of the magnetic field, which is driving the
accretion flow in the first place.  The integrated magnetic energy of
F1 always exceeds that of F1r.  It is 5\% greater through orbit 4.5,
climbs to 20\% at orbit 5, and drops back to 5\% by the
simulation's end.  During the last two orbits, the total magnetic
energy ranges from about 50\% to 70\% of the thermal energy.  Finite
resistivity decreases the turbulence levels and therefore the accretion
rate as well.  The result is that at late times, the internal energy in the
resistive run can be smaller in isolated patches, while remaining
largely unchanged in others.  Figure~\ref{temperature} shows the
central temperature as a function of radius at the end of F1 and F1r.
The temperature is roughly proportional to $R^{-3/2}$ within the
accreting disk for both models.

\section{The Generic NRAF Model}

\subsection {The Three Component Model}

In this section
we assemble into a simple model those features suggested by our
simulation that we expect will prove to be enduring and representative
of radiatively inefficient accretion flows more generally.  This model
has the virtue of limited flexibility, is falsifiable,
and is grounded in uncontroversial dynamics.  Whether it enjoys
subsequent support or must ultimately be abandoned, it will have
served a useful role.  It is schematically illustrated in
Figure~\ref{schematic}.

The structure emerges as follows.  Differentially rotating magnetized
gas is subject to the MRI.  The MRI produces a turbulent $R\phi$
Maxwell stress, and the angular momentum transport results in
accretion.  (A somewhat smaller $R\phi$ Reynolds stress component is
also present.) The sizable outward transport of angular momentum
rapidly changes the flow profile to near Keplerian, at which point the
supporting pressure gradients are small.  The global energy minimum
state separates the matter in the disk's inner region from the angular
momentum in its outer regions, as in the classical disk evolution
envisioned by Lynden-Bell \& Pringle (1974). The bulk of the gas
is decidedly disk-like: {\em whether it radiates or not,} the main body
of the accretion flow is nearly Keplerian.  Absent radiation, the disk
is simply hot, vertically thickened, and radial pressure gradients are
minimized.

The disk temperature increases rapidly inward, with $T\propto
R^{-3/2}$.  The vertical structure is dynamic, with gas lofting
away from the equator, as pressure and centrifugal accelerations
drive the gas out.  This produces the hot, dynamic coronal envelope
surrounding the disk.  Disk-generated magnetic field also rises into
the corona, where the resulting magnetic pressure significantly exceeds
the gas pressure.

The disk remains vertically thick as it accretes, making an encounter
with the centrifugal barrier inevitable, despite the loss of angular
momentum.  This is the centrifugal funnel wall.  It is present because
gravity weakens with increasing vertical distance from the central hole
while the centrifugal force remains unchanged.  Just outside the
marginally stable orbit, a small hot torus of gas accumulates.  The
specific angular momentum in the torus is slightly greater that of the
marginally stable orbit.  Hot gas, pressed up against the funnel wall,
accelerates up along this centrifugal barrier, and is held against it
by the magnetic pressure of the surrounding corona.  This is the
magnetically-confined jet, but note: the magnetic confinement is from
the outside medium!  The jet apparently is stable.

The size of the inner torus varies depending on the jet flux, the
accretion into the hole, and the rate at which matter is supplied from
the Keplerian disk.  It is highly variable.  The final
accretion into the black hole takes place {\em only} through the
opening in the funnel wall at the equator, like threading a needle.  A
high resolution torus simulation (Hawley \& Krolik 2001) found that the
magnetic stress can remain large down to and beyond the marginally
stable orbit.  This effect might increase $\dot M$ into the hole, but
the present simulation is not sufficiently well-resolved to address
this.

To summarize:  the combination of gravity, radiative inefficiency, 
angular momentum, and MHD turbulence found in black hole accretion leads
to a three component flow structure---a hot Keplerian disk, an extended
corona, and a jet-like central outflow.

\subsection{Application to Sgr A*}

An important application of ADAF-type systems has been to the source
Sgr A* at the Galactic center.  The properties of the black hole system
at the Galactic center are reviewed by  Melia \& Falcke (2001).
Compelling dynamical evidence suggests the presence of a massive
black hole of $\approx 2.6\times 10^{6}\ M_\odot$.  Observations in
X-ray and radio bands reveal a luminosity substantially below
Eddington, making this system a prime candidate as an archetype
low-radiative-efficiency accretion flow.  Recent {\it Chandra}
observations find a luminosity in 2-10 keV X-rays of $\approx 2\times
10^{33}$ erg s${}^{-1}$, and also an X-ray flare rapidly rising to a
level about 45 times as large, lasting for only $\sim 10^4$s (Baganoff
et al.~2001), indicating that the flare must originate near the black
hole.

Many aspects of spectral models for Sgr A* follow from the simple
scaling laws of black hole accretion and will be present in any model,
regardless of the detailed dynamics.
For example, Quataert \& Narayan (1999) demonstrate the impressive
range of spectra that may be generated with the adjustment of a few
free parameters:  the ratio of electron to ion temperature, the
magnetic pressure, the run of density with radius, and the accretion
rate.  Our knowledge of the flow is not yet sufficient to tightly
constrain these parameters.  As the underlying dynamical models become
more sophisticated, the spectral models should become more constrained
as well.

Although our simulation lacks a formal treatment of the energetics
necessary for a detailed application to Sgr A*, it is possible to look
at some radiative properties of the computed flow and compare them with
other, more detailed spectral predictions.  The aim is to illustrate
how the dynamical structures revealed in the simulation can be compared
with current spectral models for Sgr A*.

We must first translate between computational and  physical units, as
discussed in \S2.3.  The black hole mass is $2.6\times 10^6 M_\odot$,
which gives a Schwarzschild radius of $7.8\times 10^{11}$~cm.  To keep
things as general as possible, let the code value $n=1$ be equal to a
physical value $n_o {\rm cm}^{-3}$.  With this parameterization, the
initial torus mass is $6.8\times 10^{17}n_o$~gm.  The average accretion
rate from the inner edge of the initial torus is roughly 2\% of the
torus mass per orbit (see fig.~\ref{mdot}); in physical units this
becomes $6\times 10^{10}n_o$~gm~s${}^{-1}$, or $1.6\times 10^{-13} n_o
\dot M_{Edd}$.  The average accretion rate into the central hole is
about a factor of 10 smaller than this.

The accretion rate for Sgr A* is uncertain.  Coker \& Melia
(1997) estimate a rate of $10^{22}$~gm~s${}^{-1} = 0.03 \dot M_{Edd}$
from Bondi-Hoyle accretion of winds from nearby stars.  Reconciling
this  accretion rate estimate with the low X-ray luminosity is a
problem, however.  Quataert, Narayan, \& Reid (1999) argue that the low
luminosity requires that the accretion rate at large radius be
substantially sub-Eddington.  The best fit spectral model of Melia,
Liu, \& Coker (2001) has an accretion rate into the central hole of
$10^{16}$~gm~s${}^{-1}$.

Figure~\ref{temperature} shows the run of temperatures along the
equator in the model at the end time.  This is a single fluid
calculation with a simple equation of state, i.e., we assume that the
electron and ion temperatures are the same.  
We estimate the total bremsstrahlung emissivity using an
approximate form of the relativistic ion-electron bremsstrahlung
formula (Svensson 1982)
\begin{equation}
\Lambda = 1.2\times 10^{-22} n^2 \Theta \ln\left(\Theta+1.5\right)
{\rm ergs}\, {\rm s}^{-1}\, {\rm cm}^{-3}
\end{equation}
where $\Theta = kT/m_ec^2$, and for simplicity we use $Z=1$ and
$n_e=n_i$.  The emission is calculated for each grid zone the total is
obtained by summing up over the entire computational volume.  The total
bremsstrahlung
luminosity on the grid at the end of the simulation is $3\times
10^{21} n_o^2$~erg~s${}^{-1}$.  The inner torus dominates the total
emission.  The total thermal energy is $\sim 10^{37}n_o$~ergs, 
so the cooling time is $\sim 3\times 10^{16} n_o^{-1}$~s.  The
simulation time is $\sim 10^6$~s, so the cooling time is
substantially longer than the dynamical time for $n_o < 10^{10}{\rm
cm}^{_3}$, or $\dot M \le 10^{-3} \dot M_{Edd}$.   We note
that the formal electron-ion equilibration time from Coulomb collisions
(Spitzer 1962) is long compared to the flow time for
$n_o \sim 10^8\,{\rm cm}^{-3}$.

We can develop a qualitative sense of the bremsstrahlung spectrum by
computing the nonrelativistic value $\propto n^2 T^{-1/2} e^{-h\nu/kT}$ for
each grid zone.  At the end of the simulation, the bremsstrahlung
emission from the hot Keplerian disk inside of $R=100r_g$ peaks at a
few times $10^{19}$Hz.  The inner torus contributes the bulk of the
highest frequency emission, peaking at $10^{21}$Hz.  The coronal gas
outside of one scale height from the disk emits over a broad range of
frequency, but because the density is lower, the total emissivity is
about a factor of 10 below that of the Keplerian disk at lower
frequencies, and 2.5 orders of magnitude below at the highest
energies.  Our model is constrained by the observed low X-ray flux in
the same way as all accretion models:  the net accretion rate must be
low.  We are aided in achieving this by the coronal backflow, which
permits the escape of matter with little additional bremsstrahlung
emission, while reducing the hot gas density
near the hole.

In theoretical spectral models, the radio and submillimeter emission
arises in the innermost regions of the flow from synchrotron emission
and Compton scattering.   Synchrotron emission is governed by the
electron number and energy densities, and the strength of the magnetic
field.  In some analyses, the field strength $\beta$ and electron
temperature are treated as free parameters.  In the model of Melia et
al.~(2001) $\beta$ comes directly from the connection between the field
strength and the magnetic stress that drives the accretion.  This is
closer in spirit to a direct numerical simulation in which $\beta$
emerges self-consistently.

\begin{figure}
\plotone{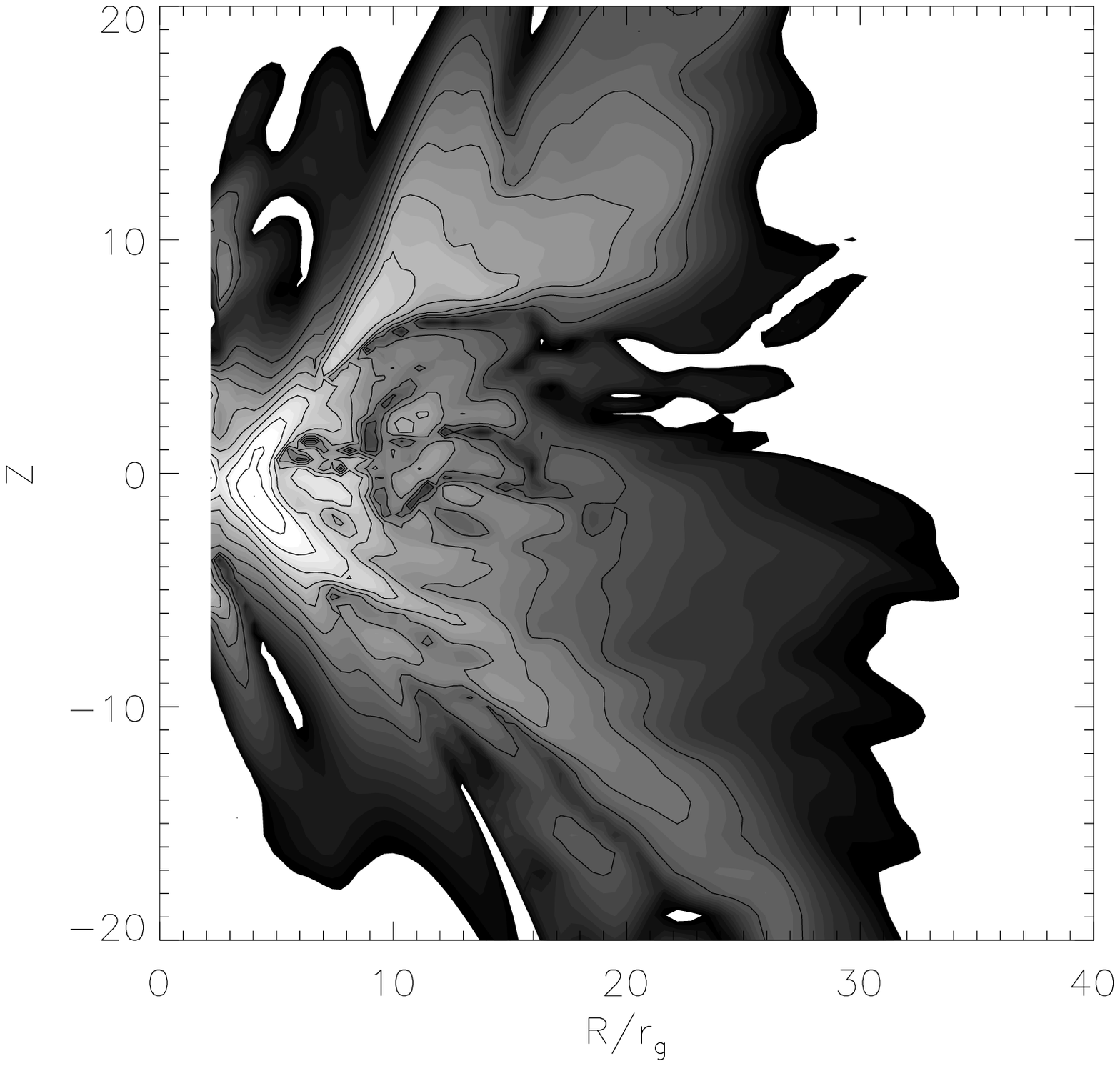}
\caption{Contour map and greyscale of peak frequencies for 
synchrotron emission at the end of simulation F1.  
The highest frequencies emerge from the
inner edge of the inner torus, 
The contours are equally spaced in log frequency;
the peak frequency is $2.5\times 10^{11}$~Hz.}
The unit number density used is $n_o = 10^8$~cm${}^{-3}$.  
\label{syncpeak}
\end{figure}

We shall limit our analysis of the sub-mm excess to a calculation of the
spatial distribution of the peak synchrotron frequency
(Rybicki \& Lightman 1979)
\begin{equation}\label{nusubc} 
\nu_c = 10^{-20} (n_e T^5/\beta)^{1/2}\ {\rm Hz},  
\end{equation} 
postponing a detailed spectral analysis to a future paper.
Figure~\ref{syncpeak} shows that the region of high temperature flow in
the inner torus can account for the observed sub-mm emission.  The
highest peak frequencies are $\sim 2.5\times 10^{11}
(n_o/10^8)^{1/2}$~Hz, and they are found at the inner edge of the disk,
where gas is compressed against the centrifugal barrier.  The region
immediately surrounding the inner disk torus is also an emitting
region.  The value of $\beta$ is $\sim 10$ at the inner edge of the
torus and $\le 1$ in the surrounding area.  A simple estimate for the
synchrotron cooling time (Spitzer 1962) is $8\times 10^5 \beta
(10^8/n_o)$~s, which is longer than the flow time.

\begin{figure}
\plotone{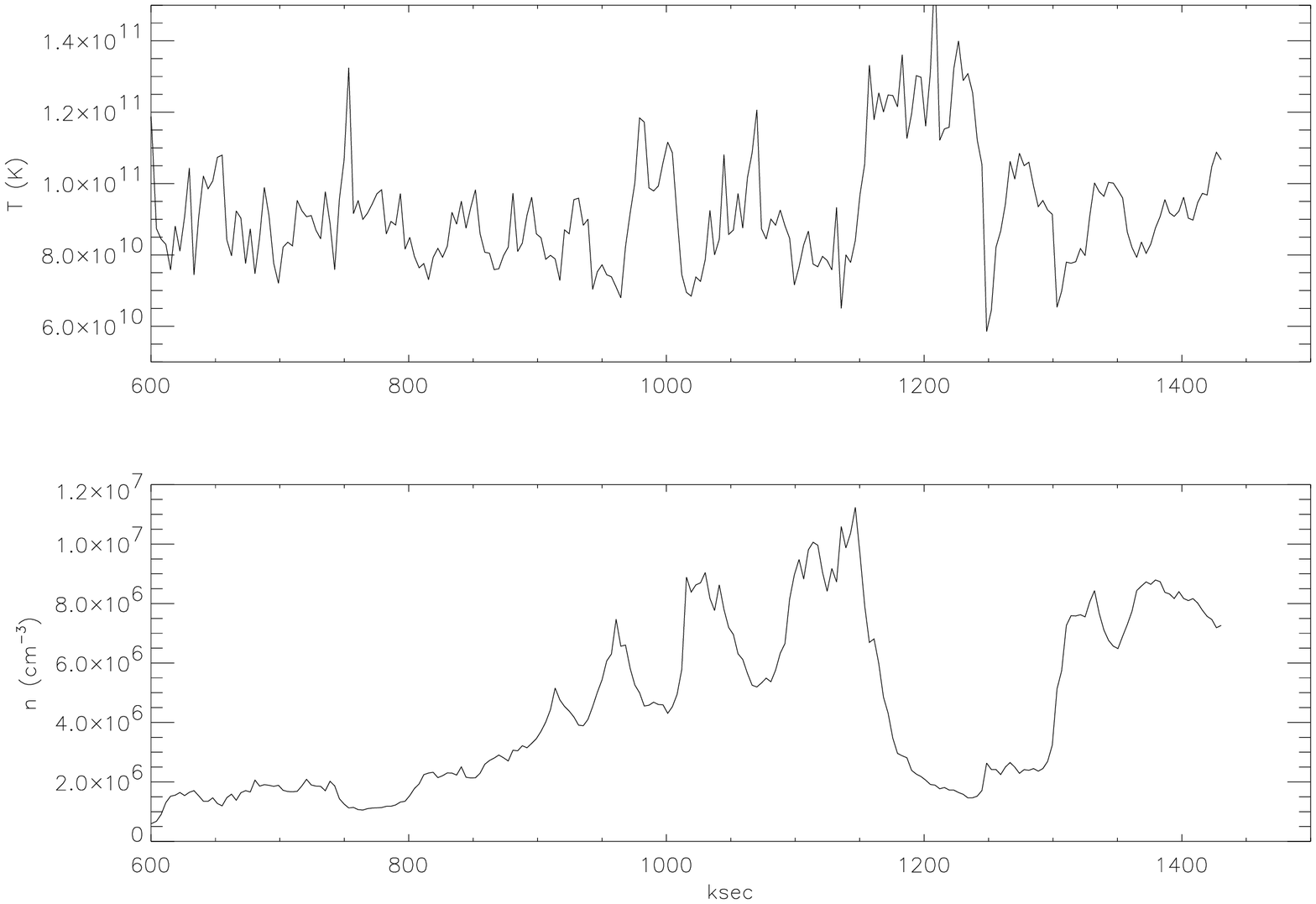}
\caption{Plot of time history of the
height- and azimuthally-averaged temperature (top) and
number density (bottom) in the inner torus at $r=4r_g$.
The numerical values are scaled to the parameters of Sgr A*, with an
assumed unit number density $n_o=10^8$ cm${}^{-3}$.  Time is
given in kiloseconds from the beginning of the simulation.}
\label{timevar}
\end{figure}

It is interesting to compare this result with the spectral model of
Melia et al. (2001).  Their best-fit model is an accretion flow with
$\dot M = 10^{16}$~gm~s${}^{-1}$, from which they compute the emission
emerging from a region inside of $5r_g$.  The gas has a temperature
$\sim 10^{11}$K, number densities $\sim 10^7$~cm${}^{-3}$ and $\beta
\approx 30$.  In our simulation the temperature in the inner torus is
$4\times 10^{10}$--$10^{11}$K, $\beta\sim 10$, and the maximum number
density along the equator lies between $n\sim n_o$ and $10\,n_o\,{\rm
cm}^{-3}$.  The inner torus is time varying over timescales of
tens of hours, with fluctuations in temperature of about 50\%, and
vertically-averaged number density $\langle n\rangle$ by about a factor
of 5 (fig.~\ref{timevar}).  The dynamics are consistent with
significant emission variability.  This is encouraging, but
given the very simple treatment of energy
in our simulation, at present it is only suggestive.

It is interesting to note that key features in our simulation
have been suggested independently by others.  For example, the jet
outflow from the inner torus may also be a source of emission, as in
the jet model of Sgr A* by Falcke \& Markoff (2000).  The observational
evidence for coronal outflows in low luminosity black hole sources
has recently emphasized out by Merloni \& Fabian (2001).
Combining the more sophisticated spectral treatments of these models
with the dynamics observed here is an obvious next step.

\section{Conclusions}

In this paper, we have examined two global MHD simulations of a
magnetized nonradiative optically thin accretion flow onto a black
hole.  Thermal conduction is not included, and the gas obeys a $\gamma
= 5/3$ adiabatic equation of state.  The initial state is a constant
angular momentum torus located at 100 Schwarzschild radii from the
hole.  Such flow is unstable according to the generalized adiabatic
criteria of Balbus (1995), becomes fully turbulent, and in the process
transports angular momentum and energy outward and greatly enhanced
rates.  Several distinctive properties characterize the resulting
flow.

Three principal flow structures are apparent: a hot Keplerian
disk that extends down to the last marginally stable orbit, a coronal
envelope, and a jet along the centrifugal barrier, or funnel wall.  

Although the flow is nonradiative and hot, it remains centrifugally
supported.  The density-weighted specific angular momentum distribution
is nearly Keplerian, $l \propto r^{1/2}$.  There are only small
sub-Keplerian departures from this.  The near-Keplerian distribution
arises naturally from the vigorous MRI-induced turbulence, which
transports angular momentum outwards until radial pressure gradients
are minimal.  No circumstances have been observed in any simulations
done to date where a significant non-Keplerian angular momentum
distribution (e.g.  constant=$l$) could be sustained over a large 
radial extent in the face of the MRI transport.

The only location where radial pressure support becomes somewhat
important is in the innermost region of the flow where a hot thickened
toroidal structure forms.  This is constantly created and destroyed.
Such a hot inner torus may be the origin of the submillimeter excess
associated with Sgr A$^*$.

The entropy and temperature increase inwards.  In the simulation the
entropy is increased by shock heating (artificial viscosity), and
resistivity.  Despite this entropy gradient, convection appears to play
no role in the dynamics of the flow, whose turbulent properties are
determined entirely by the MRI.  Marginal stability to the classical
H\o iland criteria is not achieved, nor could it be achieved as a
matter of principle (Hawley et al. 2001).  Thus, although little of
the mass that accretes at large radius makes it into the hole (in
contrast to an ADAF), this is not a CDAF.  Mass and energy are
carried off by an outflow, more in keeping with the outline of the
ADIOS model.

The flow is highly dynamic, inherently multidimensional, and
time-variable on all scales.  The flow reacts locally to the MRI,
which, because it is a fast growing instability, changes on timescales
shorter than the local orbital time.  Note that even if there exists a
time-averaged description of the global flow that is relatively
time-steady, this need not correspond to an analytic global
equilibrium.

Because these features follow directly from the self-consistent MHD
dynamics, we believe that they will be generic to nonradiative
accretion flows.  Longer evolutions would go farther in
establishing that the simulation results have lost memory of the
initial conditions.  Also, models with a greater radial domain would
allow the inflow to proceed over several decades in distance and to
liberate greater amounts of gravitational energy.  Higher resolution is
needed in the inner region to capture the details of the disk
boundary and the flow past the marginally stable orbit.  

Our data sets in principle readily lend themselves to observational
modeling.  A very preliminary application to Sgr A* relies on low
accretion rates into the hole and emission from a hot, time-varying
inner region.  A detailed follow-up investigation on the radiative
properties of our simulated flows is currently being pursued.

\acknowledgements{We acknowledge support under NSF grant AST-0070979,
and NASA grant NAG5-9266.  The simulations were carried out on the IBM
Bluehorizon system of the San Diego Supercomputer Center. 
We thank Julian Krolik and James Stone for useful discussions related
to this work.}

\end{document}